\documentclass[aps,twocolumn]{revtex4}
\usepackage{graphics}
\usepackage{epsfig}
\usepackage{color}

\newcommand{\ket}[1]{\left | #1 \right \rangle}

\newcommand{\be}{\begin{eqnarray}}
\newcommand{\ee}{\end{eqnarray}}

\begin{document}
\title{Relative phase fluctuations of two coupled one-dimensional
condensates}
\author{Nicholas K Whitlock$^{(1)}$ and Isabelle Bouchoule$^{(2)}$}
\affiliation{(1) : Department of Physics, University of Strathclyde, Glasgow
G4
0NG, UK, \\
(2) : Institut d'Optique, 91 403 ORSAY Cedex, France}

\begin{abstract}
We study the relative phase fluctuations of two one-dimensional condensates
coupled along their whole extension with a local single-atom interaction.
The  
thermal
equilibrium is defined by the competition  between independent longitudinal 
thermally excited phase fluctuations and the coupling between the 
condensates which locally favors identical phase. 
We compute the relative phase fluctuations and their 
correlation length as a function of the temperature and the strength of the 
coupling.
\end{abstract}

\maketitle

\section{Introduction}
 Recently, longitudinal phase fluctuations in very elongated 
Bose-Einstein condensates
have been observed experimentally \cite{Dettmer2002,Richard2002}.
 Such phase fluctuations are characteristic of one-dimensional (1D) Bose
gases and appear in the small interaction regime where  
$\rho \gg \sqrt{m\rho g}/\hbar$, $\rho$ being the 
linear density of atoms, $g$ the interparticle interaction 
between atoms and $m$ their mass. The opposite limit, 
called the Tonks regime \cite{tonks}, 
where strong correlations between
atoms appear is not investigated in this paper.
For 1D Bose gases, at temperatures $T$ much smaller than 
$T_{\rho}=\hbar\rho\sqrt{\rho g/m}/k_B$, fluctuations
of density are suppressed and one has a 
quasi-condensate \cite{theseDima,Phasefluctu_Petrov,lowdimension_stoof,
Phasefluctu_stoof,Quasibec_Castin}. 
However fluctuations of phase, given by
$$\langle (\theta(0) -\theta(r))^2\rangle \simeq \frac{\sqrt{m\rho
g}}{\pi\hbar\rho}
\ln(\sqrt{mg\rho}\,r/\hbar)+\frac{mk_B T r}{\hbar^2\rho}$$
are still present \cite{theseDima}. 
The logarithmic zero temperature term is negligible
when using normal experimental parameters and phase fluctuations are
produced
by the thermal population of collective modes. 

 In this paper we are interested in the case of two elongated condensates 
coupled along their whole extension by a single-atom 
interaction which enables local 
transfer of atoms from one condensate to the other. 
Such a situation could be achieved using a Raman or RF coupling between
different
internal states\cite{Matt98}. It could also model the case of condensates in two very 
elongated traps coupled by a tunnelling effect.
 The physics of two coupled condensates, 
which contains the Josephson oscillations,
has been studied in a two-mode model in \cite{Smerzi97_Josephson,
Java99_Josephson,Stringari2001_Josephson}. 
In particular the many body ground state \cite{Java99_Josephson} 
and the thermal equilibrium state \cite{Stringari2001_Josephson} 
have been computed. 
 Behind the two-mode model the excitation spectrum of 
two-component condensates coupled
by a local single-atom coupling has been calculated 
using the Bogoliubov theory in \cite{BogoinmulticomponentBEC_Meystre}.
 In the case of two elongated condensates two effects act in opposite
directions. 
Longitudinal phase fluctuations in each condensate tend to smear out the 
relative phase between the two condensates, while the coupling between 
the condensates
energetically favors the case of identical local relative phase. 
The goal of this paper is to determine the relative 
phase of the two condensates at 
thermal equilibrium as a function of the strength of the coupling.

\section{Formalism}
\begin{figure}[ht]
\includegraphics{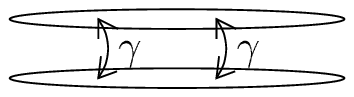}
\caption{\it
Situation studied in this article. Two elongated condensates are
coupled by a interaction which enables local transfer of atoms from
one condensate to the other.  }
\end{figure}

We are interested in pure 1D condensates where the temperature, the
interaction
energy and the coupling strength $\gamma$ are all much smaller than the
transverse
confinement energy. Thus the Hamiltonian is written
\be
H &=& \int dz\left\{\frac{-\hbar^2}{2m}
\left[\psi^{\dag}_a(z)\frac{\partial^2}{\partial z^2}\psi_a(z) +
\psi^{\dag}_b(z)\frac{\partial^2}{\partial
z^2}\psi_b(z)\right]\right.\nonumber\\
&& + \left[U(z)-\mu\right]\left[\psi^{\dag}_a(z)\psi_a(z) +
\psi^{\dag}_b(z)\psi_b(z)\right]\nonumber\\
&& +
\frac{g}{2}\left[\psi^{\dag}_a(z)\psi^{\dag}_a(z)\psi_a(z)\psi_a(z)
+ \psi^{\dag}_b(z)\psi^{\dag}_b(z)\psi_b(z)\psi_b(z)\right]
\nonumber \\&&\left.- \gamma\left[\psi^{\dag}_a(z)\psi_b(z) +
\psi^{\dag}_b(z)\psi_a(z)\right]\right\},
\ee
where $\psi_{a,b}$ are the boson annihilation operators for the
condensates labelled $a$ and $b$, $U(z)$ is the trapping
potential and $\mu$ is the chemical potential. 
Assuming that the size of the transverse ground state   
 $a_\perp=\sqrt{2\hbar/m\omega_{\perp}}$
is much larger than the s-wave scattering length $a$, the effective 
coupling constant is simply  
$g=\frac{2\hbar^2}{m}\frac{2a}{a_{\perp}^2}$.

Following calculations made for 1D condensates
\cite{theseDima,Quasibec_Castin} 
we expand the field operators in terms of their density $\rho$ and
phase $\theta$ as
\be
\psi_{a,b}(z) = e^{i\theta_{a,b}(z)}\sqrt{\rho_{a,b}(z)}.
\ee
The Hermitian density and phase  operators  obey 
$
 \left[\rho_i(z),\theta_j(z')\right] = i\delta(z-z')\delta_{i,j}
$ \cite{note.opphase}.

 As we are interested in temperatures small enough to be in the 
quasi-condensate regime,
density fluctuations are small and we write
\begin{equation}
\rho_{a,b}(z) = \rho_0(z) + \delta\rho_{a,b}(z),
\end{equation}
where $\frac{\delta\rho_{a,b}}{\rho_0}\ll 1$ and
$\rho_0$ satisfies the Gross-Pitaevskii equation
modified by taking
$\mu\rightarrow\mu + \gamma$:
\begin{equation}
\left[\frac{-\hbar^2}{2m}\Delta + U(z) - (\mu + \gamma) +
g_0\rho_0\right]\sqrt{\rho_0} = 0.
\end{equation}
We also assume that the phase difference between the
condensates at a given position is small
\begin{equation}
|\Delta\theta(z)| = |\theta_a(z)-\theta_b(z)|\ll1.
\end{equation}
The Heisenberg evolution 
equations for $\rho_{a,b}$ and $\theta_{a,b}$ are
developed to first order in $\delta \rho_{a,b}$,  
$\nabla \theta_{a,b}$ and $\Delta \theta$ and we obtain
\begin{eqnarray}
\hbar\partial_t\theta_{a,b} &=& 
-\frac{1}{2\sqrt{\rho_0}}\left [
-\frac{\hbar^2}{2m}\Delta+U+3g\rho_0- 
\mu
\right ]\left ( \frac{\delta\rho_{a,b}}{\sqrt{\rho_0}}\right )\nonumber\\
 &&   
+\gamma \frac{\delta\rho_{b,a}}{2\rho_0}
\\
\hbar\partial_t\delta\rho_{a,b} &=&
2\sqrt{\rho_0}\left [
-\frac{\hbar^2}{2m}\Delta+U+g\rho_0-\mu
\right]\left ( \theta_{a,b}\sqrt{\rho_0}\right )\nonumber\\
   &&-2\gamma\rho_0 \theta_{b,a}.
\end{eqnarray}
The first terms on the right hand side are identical to those 
for a single condensate and the second terms couple the two condensates.
We perform a canonical transformation to the bosonic operators
\be
B_{a,b} = \frac{\delta\rho_{a,b}}{2\sqrt{\rho_0}} +
i\sqrt{\rho_0}\theta_{a,b},
\ee
which evolve according to 
\be
i\hbar\partial_t\left(\begin{array}{c}
B_a\\B_a^{\dag}\\B_b\\B_b^{\dag}\end{array}\right) =
\left(\begin{array}{cc}\mathcal{L}_{GP}(\mu) & \Gamma\\
\Gamma & \mathcal{L}_{GP}(\mu)\end{array}\right) \left(\begin{array}{c}
B_a\\B_a^{\dag}\\B_b\\B_b^{\dag}\end{array}\right),
\label{eq.evolB}
\ee
where we have introduced the operators
\begin{eqnarray}
\mathcal{L}_{GP}(\alpha)\!\! & = &\!\!\! \left(\! \begin{array}{cc}
\frac{-\hbar^2}{2m}\Delta + U\! -
\alpha\! + 2g\rho_0 & g\rho_0\\
-g\rho_0 & \frac{\hbar^2}{2m}\Delta - U\! + \alpha\! -
2g\rho_0 \end{array} \! \right)\nonumber\\
\Gamma & = & \left(\begin{array}{cc} -\gamma & 0\\
0 & \gamma\end{array}\right).
\end{eqnarray}

Such an evolution is the same as the one given by the standard Bogoliubov 
theory and
we recover indeed the same result as that of
\cite{BogoinmulticomponentBEC_Meystre}.  
As the matrix in Eq.(\ref{eq.evolB}) is invariant by exchange of $a$ and
$b$, 
eigenvectors may be split in two families: the symmetric 
eigenvectors invariant by exchange of $a$ and $b$ and the 
antisymmetric eigenvectors which are multiplied by -1
by exchange of $a$ and $b$.
The eigenvalue equations are thus reduced to two $2\times2$ matrix
equations. For the symmetric family the eigenvalue equation
becomes 
\begin{equation}
\mathcal{L}_{GP}(\mu+\gamma)\left(\begin{array}{c}u_{sk}
\\v_{sk}\end{array}\right)  =\epsilon_{sk}
\left(\begin{array}{c}u_{sk}
\\v_{sk}\end{array}\right)
\label{Bogosym}
\end{equation}
and for the antisymmetric family it becomes
\begin{equation}
\mathcal{L}_{GP}(\mu-\gamma)\left(\begin{array}{c}u_{nk}
\\v_{nk}\end{array}\right)  =\epsilon_{nk}\left(\begin{array}{c}u_{nk}
\\v_{nk}\end{array}\right).
\label{eq.Bogoantisym}
\end{equation}
As for the standard Bogoliubov theory the Hamiltonian is then written, up
to a real factor, as a sum of independent bosonic excitations
\be
H_2=\sum_k \epsilon_{sk} b_{sk}^{\dag} b_{sk}+\sum_k \epsilon_{nk}
b_{nk}^{\dag}
b_{nk}
\ee
and the $B$ operator is written
\be
B_{a,b}=\sum_{k}( b_{sk} u_{sk}+b_{sk}^{\dag} v_{sk}^*)\pm
\sum_{k}( b_{nk} u_{nk}+b_{nk}^{\dag} v_{nk}^*)
\ee
 where the sums are done only on the eigenvectors normalized to
$\int dz(|u_{k}|^2-|v_{k}|^2)=1/2$.

We are interested in the  correlation function of the phase difference 
$\Delta\theta$ which is written,
after commuting the $B$ operators to normal order, 
\be
\langle\Delta\theta(z)\Delta\theta(z')\rangle =
\langle:\Delta\theta(z)\Delta\theta(z'):\rangle +
\frac{\delta(z-z')}{2\rho_0}.
\ee
The second term accounts for the phase fluctuations in a
coherent state with linear density $\rho_0$ for $a$ and $b$.
We are not interested in this term, and thus we will
consider only the normal ordered expectation value. If we expand
this in terms of the $b$ operators and consider thermal
equilibrium where no correlations between different excitations exist,
we obtain
\begin{eqnarray}
\langle:\!\Delta\theta(z)\Delta\theta(z')\!:\rangle &\!=\!&
\frac{1}{\rho_0}\sum_{k}\left\{
\langle\hat{b}_{nk}^{\dag} \hat{b}_{nk}\rangle (f_{nk}^-
{f_{nk}^-}^*{}' 
+ f_{nk}^-{}'  f_{nk}^-{}^*)\right.\nonumber\\
% &\left. - v_{nk}^*\left(u_{nk}{}' - v_{nk}{}'\right) -
%v_{nk}\left(u_{nk}^*{}' - v_{nk}^*{}'\right)\right\}
 &&\left. - v_{nk}^*f_{nk}^-{}' -
v_{nk}{f_{nk}^-}^*{}'\right\}
\label{eq.correlation}
\end{eqnarray}
where the prime means that we evaluate the function at $z'$ and
$f_{nk}^-=u_{nk}-v_{nk}$. As expected only the antisymmetric modes
contribute
because we are interested in phase difference.
% This expression, which gives the relative phase fluctuations once the 
% modified Bogoliubov spectrum of Eq.(\ref{eq.Bogoantisym}) has been
% calculated,
% is the main result of the paper.
% In the following we will give explicit results in the case of an homogeneous
% gas. 
This expression gives 
the relative phase fluctuations once the 
modified Bogoliubov spectrum of Eq.(\ref{eq.Bogoantisym}) has been
calculated. 
In the following we will give explicit results in the case of an homogeneous
 gas.

\section{Results for homogeneous condensates}
 We now consider an homogeneous gas with periodic boundary conditions 
in a box of size $L$. The potential $U$ then vanishes and the
Gross-Pitaevskii
equation gives 
\be
\mu=g\rho_0-\gamma.
\ee
The Bogoliubov function can be looked for in the form
\be
u_{sk} &=& (2L)^{-\frac{1}{2}}\exp(ikz)U_{sk}\nonumber\\
v_{sk} &=& (2L)^{-\frac{1}{2}}\exp(ikz)V_{sk}
\label{eq.planewave}
\ee
where $|U_{sk}|^2-|V_{sk}|^2=1$ 
and similarly for the antisymmetric modes.
The Bogoliubov eigenvalue equation for the symmetric branch then 
reduces to the standard Bogoliubov equation
\be
\left(\begin{array}{cc} \frac{\hbar^2k^2}{2m} + g\rho_0 & g\rho_0\\
-g\rho_0 & -\left(\frac{\hbar^2k^2}{2m} + g\rho_0\right)
\end{array}\right)\!\!\left(\begin{array}{c}U_{sk}
\\V_{sk}\end{array}\right)\!\!
=\epsilon_{sk}\!\!\left(\!\!\begin{array}{c}U_{sk}
\\V_{sk}\end{array}\!\!\right)
\ee
 whose spectrum and eigenvectors are well known.
For the antisymmetric case the eigenvalue equation becomes
\be
\left(\!\!\begin{array}{cc} \frac{\hbar^2k^2}{2m}\! +\! g\rho_0\! +\!
2\gamma &
g\rho_0\\
-g\rho_0 & \frac{-\hbar^2k^2}{2m}\! -\! g\rho_0\! -\!
2\gamma
\end{array}\!\!\right)\!\!\left(\!\!\begin{array}{c}U_{nk}
\\V_{nk}\end{array}\!\!\right)\!\!
=\!\epsilon_{nk}\!\left(\!\!\begin{array}{c}U_{nk}
\\V_{nk}\end{array}\!\!\right)
\ee
which is simply the same as the symmetric case, with the
kinetic energy shifted by $2\gamma$. Thus the eigenvalues and
eigenvector components are
\begin{equation}
\begin{array}{l}
\epsilon_{nk} = \left[\left(\frac{\hbar^2k^2}{2m}\! +\!
2\gamma\right)\left(\frac{\hbar^2k^2}{2m}\! +\! 2\gamma\! +\!
2g\rho_0\right)\right]^{\frac{1}{2}}\\
U_{nk} + V_{nk} = \left(\frac{\frac{\hbar^2k^2}{2m}\! +\!
2\gamma}{\frac{\hbar^2k^2}{2m}\! +\! 2\gamma\! +\!
2g\rho_0}\right)^{\frac{1}{4}}\\
U_{nk} - V_{nk} = \left(\frac{\frac{\hbar^2k^2}{2m} +
2\gamma}{\frac{\hbar^2k^2}{2m} + 2\gamma +
2g\rho_0}\right)^{-\frac{1}{4}}.
\end{array}
\label{eq.expUV}
\end{equation}

This two-branch spectrum was already obtained in a more
general case  in \cite{BogoinmulticomponentBEC_Meystre}.
 In the case where $\gamma \gg g\rho_0$, these excitations
are almost purely particles with $V_{nk}\ll U_{nk}$ for any $k$ and their
energy is simply $\hbar^2k^2/2m+2\gamma$ as expected for a particle in
the state $(\ket{a}-\ket{b})/\sqrt{2}$ and of momentum $k$.
 In the opposite case where $\gamma \ll g\rho_0$, three zones can 
be identified. For $k\ll 2\sqrt{m\gamma}/\hbar$ we obtain collective
excitations
with $V \simeq U$ and with energy $2\sqrt{\gamma g\rho_0}$.
For  $2\sqrt{m\gamma}/\hbar\ll k \ll 2\sqrt{mg\rho_0}/\hbar$ we still have
collective excitations with $V \simeq U$ but their energy is given by
the normal Bogoliubov dispersion law $\hbar k\sqrt{g\rho_0/m}$.
Finally for $k\gg 2\sqrt{mg\rho_0}/\hbar$ excitations are just particles 
with energy $\hbar^2k^2/2m$.

Using the plane wave expansion (\ref{eq.planewave}) and the normalization
condition
$U_{nk}^2-V_{nk}^2=1$ the correlation function 
(\ref{eq.correlation}) of the relative phase fluctuation is written 
\begin{eqnarray}
%\begin{array}{l}
&&\langle:\Delta\theta(z)\Delta\theta(z'):\rangle
=\frac{1}{2\rho_0L}\nonumber\\
&&\sum_{k}\left\{\left(U_{nk} -
V_{nk}\right)^2 \left(2n_{nk} + 1\right) - 1\right\}\cos[k(z-z')],
%\end{array}
\label{eq:phasesum}
\end{eqnarray}
where $n_{nk}=1/(e^{\epsilon_{nk}/k_B T}-1)$ 
is the occupation number for the state with energy
$\epsilon_{nk}$. Using the expression (\ref{eq.expUV}) this correlation
function can be computed numerically. In the following we analytically
compute
the phase fluctuations using some approximations.

 The terms which do not involve $n_k$ correspond to the zero temperature
contribution.
As the function $V_{nk}^2-U_{nk}V_{nk}$
is always smaller than the corresponding
function for a single condensate, the relative phase fluctuations will be
smaller than the phase fluctuations of a single condensate which implies
\be
\langle :\Delta\theta^2:\rangle < \frac{\sqrt{mg\rho_0}}{\hbar\rho_0}
\ln\left(\frac{L\sqrt{mg\rho_0}}{\hbar}\right).
\ee
The whole theory is valid only for large density so that 
$\sqrt{mg\rho_0}/(\hbar\rho_0)\ll 1$ and in the experiments accessible until
now
the size of the condensate is not large enough to produce noticeable 
phase fluctuations at zero temperature.

 Phase fluctuations are thus due to thermal excitation of the collective
modes
and
we will give a simplified expression by making  several approximations.
First we will approximate the Bose factor by
\be
n_k=\frac{k_B T}{\epsilon_{nk}}.
\ee
This is justified as 
this expression deviates in a significant way from the Bose occupation
factor 
only when $n_k$ becomes smaller than 1, ie when $\epsilon_{nk}>k_B T$, and
the contribution to phase fluctuations of those modes is small even with the
previous expression which overestimates their population. 
Let us now consider separately the case where $\gamma \gg g\rho_0$ and
the case $\gamma \ll g\rho_0$.

If $\gamma \gg g\rho_0$, then $(U_{nk} -
V_{nk})^2\simeq 1$ for all $k$ and $\epsilon_k\simeq \hbar^2k^2/2m+2\gamma$.
This gives, approximating the discrete sum by an integral, 
\begin{eqnarray}
%\begin{array}{ll}
\langle:\Delta\theta(z)\Delta\theta(z'):\rangle &=&
\frac{2k_BT}{2\pi\rho_0}\int_{-\infty}^\infty dk
\frac{\cos[k(z-z')]}{\hbar^2k^2/m+4\gamma}\\
&=&\frac{k_B T}{2\hbar\rho_0}\sqrt{\frac{m}{\gamma}}
e^{-2|z-z'|\sqrt{m\gamma}/\hbar}.
%\end{array}
\end{eqnarray}
As we consider only temperatures $k_B T\ll \hbar\rho_0 \sqrt{g\rho_0/m}$ so
that we have quasi-condensate, these phase fluctuations are always 
very small.

\begin{figure}[ht]
\begin{center}
\epsfig{figure=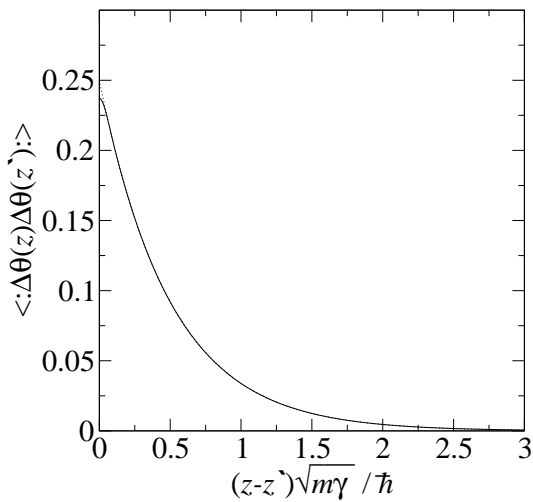,width=80mm}
\end{center}
\caption{\it Correlation function of the relative phase 
fluctuations. The solid line is the numerical calculation
of Eq.(\ref{eq:phasesum}) with $\gamma=g\rho_0/10$,
$T=\hbar\rho_0 \sqrt{\gamma}/(2\sqrt{m}k_B)$ and $L= 100
\hbar/\sqrt{mg\rho_0}$.
The dotted line is the analytical expression
Eq.(\ref{eq.correlationfonction}) which only differs from the
numerical expression at small separations.}
\label{fig.compnum}
\end{figure}

\begin{figure}[ht]
\includegraphics{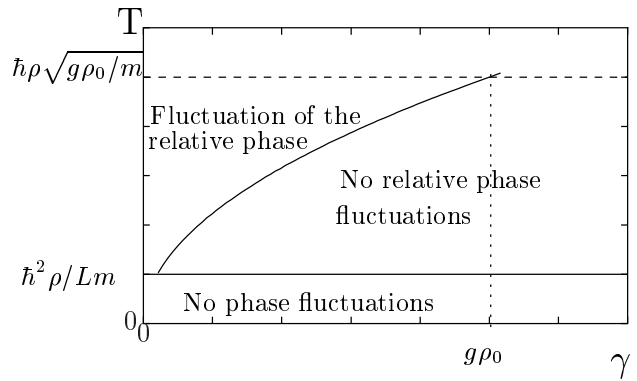}
\caption{\it Phase diagram for the fluctuations of the relative phase
between the two condensates. Only temperatures much smaller than 
$\hbar\rho_0\sqrt{g\rho_0}/(k_B\sqrt{m})$ are relevant as for larger
temperatures one
does not have a quasi-condensate anymore. For temperatures larger than 
$\hbar\rho_0/(k_BL\sqrt{m})$, each condensate has longitudinal phase
fluctuations.
Below the curve, which corresponds to Eq.(\ref{eq.Tlimite}), 
the coupling between the condensates is large enough to suppress
local relative phase fluctuations between the two condensates. 
Above this curve, there are local relative 
phase fluctuations between the two condensates.}
\label{phasediagramme}
\end{figure}

Let us now consider the case where $\gamma \ll g\rho_0$.
The modes with 
$|k|\gg k_0=\sqrt{mg\rho_0}/\hbar$ give a negligible contribution
to the phase fluctuations. Indeed for those terms 
$(U_{nk} -
V_{nk})^2\simeq 1$  and $\epsilon_k\simeq\hbar^2k^2/2m$ so that their
contribution
to the phase fluctuation is
\be
\frac{mk_B T}{\pi\hbar^2\rho_0}\int_{k_0}^\infty dk \frac{1}{k^2}=
\frac{k_B T\sqrt{m}}{\pi\hbar\rho_0\sqrt{g\rho_0}},
\ee
which is always small in the regime of quasi-condensates.
Thus only the modes $|k|\ll 2\sqrt{g\rho_0 m}/\hbar$ are considered for
which  
\be
\left(U_{nk} - V_{nk}\right)^2 \simeq 
\frac{2\sqrt{g\rho_0}}{\sqrt{\hbar^2k^2/m+4\gamma}}
\ee 
and the correlation function then becomes
\be
\langle:\Delta\theta(z)\Delta\theta(z'):\rangle \simeq
\frac{2k_BT}{\rho_0\pi}\int_0^{k_0}
\frac{dk}{\frac{\hbar^2k^2}{m} + 4\gamma}\cos[k(z-z')].
\label{eq.fluctugammasmall}
\ee
The integral can actually be extended to infinity as higher $k$ values
give negligible
contributions and we find
\be
\langle:\Delta\theta(z)\Delta\theta(z'):\rangle =
\frac{k_BT}{2\rho_0\hbar}\sqrt{\frac{m}{\gamma}}
\exp\left[\frac{-2\sqrt{m\gamma}|z-z'|}{\hbar}\right].
\label{eq.correlationfonction}
\ee
Note that this expression is the same as 
Eq.(26), which was not expected {\it a priori}.
This formula, which give the amplitude of the 
relative phase fluctuations as well as their
correlation length $1/\gamma$ is the main result of the 
paper.
It agrees well with the numerical calculation
of Eq.(\ref{eq:phasesum}) as shown in Fig.\ref{fig.compnum}.
Phase fluctuations are small only if
\be
k_B T\ll \rho_0\hbar\sqrt{\frac{\gamma}{m}}.
\label{eq.Tlimite}
\ee
Note that as we assumed small relative phase difference, 
this is also the limit of validity of  our calculation.
The phase diagram of Fig.\ref{phasediagramme} summarizes
the previous results.

\section{Dynamical interpretation}
The condition (\ref{eq.Tlimite}) to have small relative 
phase fluctuations 
has a dynamical interpretation which is shown very qualitatively below.
In a two-mode model of the Josephson coupling between two condensates of $N$
atoms 
it has been shown that if  $\gamma \ll N\partial \mu /\partial N$,  
then the Josephson oscillation frequency 
is \cite{Java99_Josephson,Stringari2001_Josephson,Smerzi97_Josephson}
\begin{equation}
\omega_J=\frac{2}{\hbar}\sqrt{\gamma N\frac{\partial\mu}{\partial N}}\simeq
\frac{\sqrt{\gamma \mu}}{\hbar}=
\frac{\sqrt{\gamma g\rho_0}}{\hbar}.
\label{eq.freqJoseph}
\end{equation}
On the other hand a single elongated condensate will experience phase 
fluctuations and the phase at a given position will evolve in time. 
If the change of the phase during a Josephson oscillation time is small,
then the Josephson coupling will ensure that the relative phase between
the two condensates remains zero: there will be no relative phase
fluctuations.
However if the change of the phase during a Josephson oscillation 
time is large, then the Josephson coupling will not have time to adjust the 
phase of one condensate with respect to the other: there will be relative
phase fluctuations of the two condensates.
 We thus have to compute the change of the local phase of a single
condensate
\begin{equation}
\langle (\theta(1/\omega_J)-\theta(0))^2\rangle,
\end{equation}
with the average corresponding to the thermal equilibrium and the coupling 
between the two condensates being ignored.
This calculation could be done rigorously by developing the operator
$\theta$  
on the collective excitation bosonic operators $b_k$ and $b_k^{\dag}$.
In the following we present a simpler argument that gives the same order of
magnitude.
We first estimate the amplitude $A_k$ of the phase 
modulation of wave vector $k$.
The energy of this phase modulation is just the kinetic energy
$N|A_k|^2\hbar^2 k^2/4m$ which in a classical field theory at thermal
equilibrium corresponds to an energy of $k_B T/2$
and thus 
\begin{equation}
|A_k|^2 \simeq \frac{2m k_B T}{\hbar^2k^2}\frac{1}{\rho_0 L}.
\end{equation}   
This is indeed the contribution of the mode $k$ to phase fluctuations 
as computed in Eq.(\ref{eq.fluctugammasmall}) if $\gamma=0$.
According to the Bogoliubov spectrum and because only modes
with $|k|\ll 2\sqrt{g\rho_0 m}/\hbar$ contribute, 
the mode $k$ evolves with the frequency 
\begin{equation}
\omega_k=\frac{k\sqrt{g\rho_0}}{\sqrt{m}} .
\end{equation}
The evolution of the phase after a Josephson oscillation time 
$t_J\simeq\hbar /\sqrt{\gamma g\rho_0}$ is then written, after averaging
over
the
independent phases of the phase modulations,
\begin{eqnarray}
%\renewcommand{\arraystretch}{2}
%\begin{array}{ll}
\langle (\theta(t_J)-\theta(0))^2\rangle &\simeq& 
\sum_k |A_k|^2 (1-\cos(\omega_k t_J))\nonumber\\
&\simeq& \frac{2mk_B T}{\hbar^2 \rho_0 2\pi}
\int_{-\infty}^\infty 
\frac{1-\cos(\sqrt{t_Jg\rho_0}k/\sqrt{m})}{k^2} dk \nonumber\\
% &\simeq \frac{k_B T \sqrt{m\rho_0 g}}{\hbar^2 \rho_0}
% \frac{\hbar}{\sqrt{\gamma \rho_0 g}}
&\simeq& \frac{k_B T\sqrt{m}}{\hbar \rho_0\sqrt{\gamma}}.
%\end{array}
\end{eqnarray} 
Small relative phase fluctuations of the two condensates occurs when this 
quantity is small and we recover the condition (\ref{eq.Tlimite}).

\section{discussion}
In conclusion we have shown that as long as the temperature is small enough
to fulfill Eq.(\ref{eq.Tlimite}), although there might exist large phase
fluctuations 
along each condensate, the local relative phase of the two condensates stays
small. 
In the opposite case there are large fluctuations of the relative phase
whose
correlation length is $l_c=\hbar/2\sqrt{m\gamma}$.
As an example let us consider the case of two Rubidium 
condensates of $10^4$ atoms elongated over
$L=200\,\mu$m, confined transversely with 
an oscillation frequency $\omega_{\perp}/2\pi=1\,$kHz and coupled using 
$\gamma=50\,$Hz. The phase of each condensate changes by about $2\pi$ from
one
end of the condensate to the other as soon as 
$T>T_\phi=\hbar^2\rho_0/(mLk_B)=1.8\,$nK. However the local relative phase
between 
the two condensates stays much smaller than 1 if 
$T \ll \hbar\rho_0 \sqrt{\gamma}/(k_B\sqrt{m})=180\,$nK.   
 The calculations made here for homogeneous condensates could be used 
to describe a trapped inhomogeneous gas {\it via} a local density
approximation
similar to that used in \cite{Gerbier2002}  
as long as both the healing length $l_h=\hbar/\sqrt{mg\rho_0}$
and the correlation length of the phase fluctuations are much smaller
than the extension of the condensate.
In the above example, $l_h=0.6\,\mu$m and $l_c= 2\,\mu$m are indeed much
smaller
than  
$L$.

To measure experimentally the relative phase fluctuations 
and their correlation length, one should perform an 
interference experiment. In the case where the two states 
are internal states, an intense $\pi/2$ pulse 
has to be applied. Measurement of the local density of atoms
in the state $\ket{a}$ and $\ket{b}$ then gives access to the local
relative phase of the two condensates.
 In the case where $\ket{a}$ and $\ket{b}$ are confined in 
the wells of a double well potential, the interference measurement is
performed {\it via} a fast release of the confining
potential followed by a time of flight long enough for the two clouds to
overlap. Indeed, the total intensity 
presents fringes in the direction orthogonal 
to $z$\cite{Ketterle-doublepuits} and, at 
a given $z$, the position
of the central fringe gives the value of the local
relative phase.

\acknowledgments
We are grateful to Alain Aspect and Stephen Barnett for
suggesting this collaboration  and for stimulating discussions. 
We thank Fabrice Gerbier and Gora Shlyapnikov for 
helpful discussions.
The authors
would like to thank the Carnegie Fund, the Overseas Research Students
Awards Scheme, the University of Strathclyde and the CNRS 
for financial support.

%\bibliography{dimreduites,Josephson,multicomponentBEC} 

 \end{document}